\documentclass[12pt,a4paper]{article}
\usepackage{epsfig}

\makeatletter
\newcommand{\LyX}{L\kern-.1667em\lower.25em\hbox{Y}\kern-.125emX\spacefactor1000}
\makeatother
\newcounter{figno}
\newenvironment{figs}{
\begin{list}{\arabic{figno}.\hfill}{\usecounter{figno} \labelwidth2em
\labelsep1em \itemsep0ex \leftmargin3em \itemindent0em}}%
{\end{list}}
\input tcilatex
\QQQ{Language}{American English}

\begin{document}

\title{Model Calculations for Copper Clusters on Gold Electrodes}
\author{M.G. Del P\'{o}polo, E. P. M. Leiva\footnote{corresponding 
author} \\
Universidad Nacional de C\'{o}rdoba\\
Unidad de Matem\'{a}tica y F\'{\i}sica, Facultad de Ciencias Qu\'{\i}micas\\
5000 C\'{o}rdoba, Argentina\\
and\\
W. Schmickler\\
Abteilung Elektrochemie, Universit\"{a}t Ulm\\
D-89069 Ulm, Germany}
\date{}
\maketitle

\begin{abstract}
Using the embedded-atom method, the structure of small copper 
clusters on Au(111) electrodes
has been investigated both by static and dynamic calculations. By 
varying the size of roughly circular clusters, the edge energy per 
atom is obtained; it agrees quite well with estimates based on 
experimental results. Small three-dimensional clusters tend to have 
the shape of a pyramid, whose sides are oriented in the directions of 
small surface energy. The presence of a cluster is found to distort the 
underlying lattice of adsorbed copper atoms.

\end{abstract}

\section{Introduction}
During the last two decades, the scanning tunneling microscope (STM) has 
become a versatile tool for the investigation of 
conducting surfaces. It has not only been used for obtaining
images  with atomic resolution, but also for the modification of the 
surface structure. In particular, experimental procedures have been 
developed for the deposition of small metal clusters on surfaces. In 
this respect electrochemical interfaces are particularly suited for a 
controlled modification of surfaces, since two potential differences 
can be varied independently: the bias between the electrode and the 
STM tip, and the electrode potential with respect to a reference 
electrode. Thus, Kolb et al. \cite{kolb1} have been been able to 
deposit small copper clusters on the surface of Au(111) electrodes by 
first depositing copper on the STM tip and subsequently bringing the 
tip in contact with the surface. This deposition can be controlled 
with great precision, and has given rise to speculations about  
possible applications in nanotechnology. Recently, these copper 
clusters have attracted further attention because they appear to be stable 
at electrode potentials at which bulk copper dissolves, an effect 
which has been attributed to quantum confinement \cite{kolb2}.

Since these copper clusters on Au(111) have become to be considered as 
a prototype for electrochemical surface modifications, we have 
performed model calculations for clusters of various sizes and shapes within the 
embedded atom method. An important characteristic of the Cu/Au(111)
system is the fact that before the formation of the cluster, which is 
performed at potentials slightly above the deposition potential for 
bulk copper,  the electrode 
is already covered with an epitaxial monolayer of copper atoms. 
Obviously, this must be taken into account if the calculations are to 
be meaningful. 

\section{Calculation method}

Since we are dealing with metallic systems comprising a large
number of atoms, we require a calculation method that allows the 
treatment of 
large systems, but without missing the important many-body properties of 
metallic binding. In this respect, the embedded-atom method (EAM) as
proposed by Daw and Baskes \cite{Daw-Baskes} appears as the natural choice.
This model for the interaction between metallic atoms in condensed phases
gives a much better description of the metallic bond than pair potentials
which do not take into account many-particle effects. 

Within the embedded-atom method the total energy 
$E_{tot}$ of an arrangement of $N$ metal particles is calculated as the sum of
individual particles energies $E_i$:
\begin{equation}
E_{\mathrm{tot}}=\sum_{i=1}^N E_i\qquad \mbox{where}\qquad 
E_i=F_i(\rho _{h,i})+\frac{1}{2} \sum_{j\neq i}V_{ij}(r_{ij})  \label{Ei}
\end{equation}
$F_i$ is the embedding function and represents the energy necessary to embed
atom $i$ into the electronic density $\rho _{h,i}$. The latter quantity is
calculated at the position of atom $i$ as the superposition of the
individual atomic electronic densities $\rho _i(r_{ij})$ of the other
particles in the arrangement:
\begin{equation} 
\rho _{h,i}=\sum_{j\neq i}\rho _i(r_{ij}) 
\end{equation}
The attractive contribution to the energy is given by the embedding function 
$F_i$, which contains the many-body effects. The repulsive
interaction between ion cores is represented as a radial pair potential 
$V_{ij}(r_{ij})$,  which takes the form of a pseudocoulombic repulsive energy: 
\begin{equation}
V_{ij}=\frac{Z_i(r_{ij})Z_j(r_{ij})}{r_{ij}} 
\end{equation}
where $Z_i(r_{ij})$ may be considered as an effective charge which depends
upon the nature of particle $i$.

We have employed two different versions of EAM potentials. On one hand, 
we have used
the original parameterization of Foiles et al \cite{Foiles} (denoted FDB),
which is based on  experimental data of pure metals and of
several alloys, so as to reproduce such parameters as  equilibrium
lattice constants, sublimation energies, bulk moduli, elastic constants,
vacancy formation energies,  and heats of solution. On the other hand, 
we have used
a potential specifically devised by \ Barrera et al.\cite{Barrera}(denoted
BTI) to describe the structure and energetics of Cu-Au alloys.

\section{Results and discussion}

\subsection{Energetics of copper islands}

The clusters generated by electrochemical STM techniques come in 
various forms and sizes. In this section we focus on flat islands 
that are only a few monolayers thick. In order to explore the 
energetics of such clusters we have performed static calculations 
within the embedded-atom scheme, and assumed a commensurate structure. 

Both the energy of  adsorption per atom in the bulk of the cluster 
and edge effects are of interest. In order to separate the two, we have 
adapted the drop model \cite{drop_model}
that is often used for the calculation of 
surface energies to the two-dimensional situation. 

In order to calculate properties for a Cu island adsorbed on a surface, we
use a two-dimensional version of the drop model. Thus, we split the energy of a 
two-dimensional cluster into a bulk and a surface contribution; for a 
circular cluster we may write:
\begin{equation}
	E = 4 \pi R^2 e_{s} + 2 \pi R e_{l}
	\label{2drop}
\end{equation}
where $R$ is the radius of the cluster, $e_{s}$ the energy per unit 
surface of the cluster, and $e_{l}$ the energy per unit boundary 
length. Rearrangement gives:
\begin{equation}
	\frac{E}{\pi R} = e_{s} + \frac{2}{R}\, e_{l}
	\label{drop2}
\end{equation}
so that the two contributions $e_{s}$ and $e_{l}$ can be determined by 
a plot of the total system energy $E$ vs. $1/R$. However, such a plot 
gave a fair scattering of data particularly for small clusters, where 
the deviations from the spherical shape are noticeable. We have 
therefore modified eq.~(\ref{2drop}) a little and written:
\begin{equation}
	E = u_{s}n_{s} + u_{l}n_{l}
	\label{2cluster}
\end{equation}
where $u_{s}$ and $u_{l}$ denote the energy per bulk atom and per
boundary atom, respectively, and $n_{s}$ and $n_{l}$ denote the 
numbers 
of bulk and of boundary atom. Rearrangement now leads to:
\begin{equation}
\frac E{n_t}=\ u_s+\left( u_l-u_s\right) \frac{n_l}{n_t}\   \label{Esys4b}
\end{equation}
where $n_t=n_s+n_l$ is the total number of atoms in the cluster.

A typical plot of $ E/n_t$ vs. $n_{l}/
n_t$  is shown in Fig.~1. The fit to a straight line is excellent, 
so that both the bulk adsorption energy $u_{s}$ and the quantity
$\Delta u_{\mathrm{step}} = \left( u_l-u_s\right)$, which is the 
energy required to form a step atom from a bulk atom, can be 
determined with good accuracy. A few results are shown if Fig.~2;
 they have  been obtained both with the BTI and with the FDB 
potentials, which here give quite similar results.

The adsorption energy $u_{s}$ per bulk atom is highest for the
 cluster that is one monolayer high, where all the copper atoms are bonded directly 
to the gold surface. For clusters that are two or three layers high, 
the adsorption energies are the same within the accuracy of the 
calculation. In contrast, the energy for step formation is lowest for 
the monolayer, indicating that the Cu-Cu bond is weakened by the 
stronger Cu-Au bond. The energy per atom required to form a double 
step is particularly large, so that clusters of pyramidal shape are 
favored. These results bear witness to the fact that in metallic 
bonding the bond energies are not additive.

Because of the circular geometry assumed for the Cu island, calculations for
step formation of figures 2a-2c correspond to an average over different
orientations on the surface. Calculations can also be performed for
different directions on the surface, as shown in figure 2d. As expected,
 the step formation energy is lower for the
formation of more compact steps, so these should predominate.

The step-formation energy is difficult to measure, but estimates can 
be obtained from the growth and the stability of copper clusters.  Xia 
et al. \cite{xia} estimate the step energy to be of the order of 0.4 
to 0.5~eV, which is quite in line with our calculations.

\subsection{Atom dynamics simulations}

In order to explore structures that are not necessarily commensurate 
with the Au(111) substrate, we performed molecular-dynamics 
simulations within the embedded-atom scheme. At first, we tested the two
different potentials employed 
in this work and performed calculations for a Cu monolayer adsorbed on Au(111).
The system consisted of a  slab of Au atoms eight layers thick, 
four of them mobile,
covered with a Cu layer. Each of these layers consisted of 196 atoms; 
cyclic   boundary conditions were imposed parallel to the surface. For 
the the interactions between the particles a cut-off radius of 7~\AA\ 
was employed.
The time step was set at 0.2~fs, and the simulations ran over 4 ns.

Although the energetic predictions of the two parameterizations
employed here are very
similar, as shown in the previous section, they result in rather 
different structures for  the adsorbed Cu monolayer. In the case of
the FDB potentials the monolayer is compressed, leading to holes and domains
with an incommensurate structure (see Fig.~3a).
In contrast, the BTI potentials, which are based on data for Cu-Au 
alloys,
predict a commensurate $(1\times 1)$ structure, which correspond to the
experimental observation in the electrochemical system (see fig.~3b).
 It is worth mentioning that although both structures show
strong differences while observed in direct space, the Fourier transforms of
the corresponding averaged atomic densities lead to a hexagonal pattern
with the same lattice parameter, with considerably more diffuse spots in the
case of the FDB potentials. Because of its better agreement with the
experimental finding, we shall use only the BTI potentials in the following
simulations. However, the tendency of the copper atoms to form 
compressed structures is real -- it is caused by the shorter lattice 
constant of bulk copper -- and will also be found in the simulations 
with the BTI potentials reported below. 

To investigate the properties of copper clusters 
adsorbed on a Au(111) surface we have performed atom dynamics
simulations with Cu clusters containing between 55 and 125 atoms with 
different starting configurations such as a cylindrical or an 
inverted pyramidal structure; typical runs lasted   between 4 ns and 10 ns. 
Since a full
statistical analysis  requires a  higher number of runs than we 
have 
undertaken so far, we shall only give here some representative results
which show the main features of the investigated system.

Figure 4 shows the atomic arrangement for a  Cu cluster that is
7-layers high. The pyramidal structure is favored by the step 
energies (see previous section) and was obtained with various 
starting configurations. Is is  instructive to consider the 
time-averaged atomic density as a function of the distance 
perpendicular to the surface (Fig.~4b).
The decreasing height of the peaks and  their decreasing area
reflect the pyramidal structure of the cluster. Note that the 
peaks become broader with increasing distance, since the atoms in the 
top layers are more mobile.  An analysis of the distance between
lattice planes yields values in the range of  2.00 to 2.02 \AA\ for the Cu
planes closer to the surface. In bulk copper the distance between
lattice planes in the (111) direction is 2.09 \AA, so the clusters are 
somewhat contracted.
This is understandable
in terms of bond order arguments: The lower coordination of the atoms at the
boundary of the layer yields a stronger binding between the layers, with a
concomitant lowering of the interlattice distance.

The presence of a cluster induces a rearrangement of the copper 
monolayer underneath. In fact,
several atoms initially belonging to the cluster are incorporated into the
monolayer during the simulations, making it more compact after equilibration. 
This effect is illustrated
in fig.~5, where a cluster is shown along with the structure of the Cu
monolayer in contact with the Au(111) surface. The presence of the cluster
strongly  distorts  the monolayer and creates dislocations.

The compression of the monolayer can be quantitatively expressed
through the average distance to the nearest neighbors.
Figure 5c shows a corresponding contour plot for the copper atoms in 
the monolayer. The data were obtained by averaging the  distance of 
each particle to its nearest neighbors, and by further averaging over 
1000 time steps. It is evident that the
presence of the cluster produces a dramatic decrease of the 
nearest-neighbor distance in the
monolayer;  the minimal values  are around 2.67 \AA , which
is somewhat larger than the value for bulk copper (2.55 \AA),
  but much smaller than the distance between Au atoms on the 
Au(111) surface (2.88 \AA ). Figure 5c also
reveals the shape of the basis of the cluster, which resembles that of a
truncated equilateral triangle. This can be understood as follows:
According to the static calculations of the previous sections, steps should
preferentially occur in the more compact $<110>$
directions. Two kinds of facets arise in these directions, which depending
on their structure are usually denoted as \{111\} and \{100\} faces,
respectively \cite{Stumpf}. These facets originate the growth of faces of
pyramids that resemble the (111) and (100) fcc faces respectively, which are
known to have different surface energies. Since the embedded atom 
method predicts, in
qualitative agreement with experiment, a lower surface energy for the (111)
surface, this is the facet that is expected to predominate, and the
pyramids should exhibit hexagonal bases, where the sides corresponding to
the \{111\} facets should be considerably larger than those of the
\{100\} facets.

The clusters generated by these simulations are stable within the time 
that can reasonably be covered in a simulation. They are 
not stable in an absolute sense since it would generally be favorable 
to spread them out into a monolayer. However, real copper clusters, if 
they are chemically pure, should also be in a metastable state, since 
the average binding energy in a cluster is always less than that of 
the bulk material. Electrochemical experiments with such clusters are 
usually performed in sulphate solutions; in the region near the 
equilibrium potential for the Cu/Cu$^{2+}$ system copper is covered by 
a layer of adsorbed sulphate ions, which may increase their stability. 

\section{Conclusions}
The embedded-atom method, which we have employed, is particularly 
suited to treat large ensembles of metal atoms; both static and 
dynamic calculations can be performed with relative ease
\cite{landman}. Since the 
interaction
parameters have been fitted to experimental data, we expect the 
method to give qualitatively correct results and to  provide good estimates
for binding and step-formation energies. We think that our results 
for copper on Au(111) are encouraging, and that the method can be 
applied to similar systems and give valuable information about the 
properties of metal clusters on metal electrodes.

Our calculations do not explain the extra stability of copper 
clusters observed by Kolb et al. \cite{kolb2}. However, our method is 
not suited to investigate quantum confinement, which these 
authors believe to be the cause of this effect.
 On the other hand, 
metal dissolution is a comparatively slow process: it involves the 
solvation of the dissolved ion, and thus requires $10^{-12} - 
10^{-11}$~s. On this timescale, the copper clusters can freely exchange 
electrons with the underlying substrate. The very fact that these 
clusters can be imaged with the STM shows that such an exchange occurs 
without a major hindrance. Therefore the explanation for this 
stability has to be sought elsewhere.
Preliminary energetic considerations
with the present calculation method, indicate that the admixture of gold to
copper clusters may improve their stability. This will be the subject 
of further investigations. 

\subsection*{Acknowledgment}
Financial support by the DAAD, the Deutsche Forschungsgemeinschaft,  and by Fundaci\'{o}n Antorchas is gratefully
acknowledged. E.P.M.L. and M.G. D.P. thank CONICET, CONICOR, Secyt UNC and
Program BID 802/OC-AR PICT N$^o$ 06-04505 for financial support.

\newpage

\section*{Figures}
\begin{figs}

\item Binding energy per atom for a Cu island adsorbed on a clean Au(111) surface
 as a function of the fraction of atoms that belong to the border of the island.
 
\item Energy per surface atom $u_s$ and energy required to generate an atom 
at a step from an atom on the surface of the island $\Delta u_{\mathrm{step}}$ 
for several
systems. FDB denotes the results obtained with the potentials of Foiles et al. \cite{Foiles}
and BTI indicates the corresponding results with the potentials of Barrera et al.\cite{Barrera}.
 An island with circular shape  was assumed in all cases.
(a) Cu island on Au(111).
(b) Cu island on a pseudomorphic layer of Cu adsorbed on Au(111).
(c) Two-atom high Cu island on a pseudomorphic layer of Cu adsorbed 
on Au(111).
(d) Step formation energy for different directions on the surface.  The calculations were
performed with the BTI potentials.
\item Final structures obtained after 4~ns in an atom-dynamics simulation  for a
Cu monolayer adsorbed on Au(111). (a) Potentials of Foiles et al. 
(FDB) \cite{Foiles};
(b) potentials of Barrera et al. (BTI) \cite{Barrera}.
\item Atomic arrangement (a),  and  Cu atom density (b) as a function of the distance 
perpendicular to the surface for a 7-layers high Cu cluster formed on a Au(111) 
surface.
\item Compression of the Cu monolayer induced by the presence of a
cluster. (a) Atomic arrangement of the cluster. (b) Structure of the Cu layer
in contact with the Au(111) surface.
(c) Contour plot of the distance between nearest neighbors in the 
Cu monolayer. The contour lines start at a distance of 2.65 \AA, are 
separated by 0.05 \AA, and end at 2.95 \AA. The geometric arrangement is shown for comparison.
\end{figs}

\end{document}